\UseRawInputEncoding
\documentclass[aps,prl,twocolumn, showpacs,longbibliography,superscriptaddress]{revtex4-1}
\usepackage{amsmath}
\usepackage{amssymb}
\usepackage{xcolor}
\usepackage{graphicx}
\usepackage{dcolumn}
\usepackage{gensymb}
\usepackage{float}
\usepackage{bm}
\usepackage{hyperref}
\usepackage[caption=false, font=normalsize, labelfont=sf, textfont=sf,position=top,justification=raggedright,singlelinecheck=false]{subfig}
\usepackage{ragged2e}

\begin{document}

\title{Single Color Center Spin Coherence revealed in Optically Detected Magnetic Resonance of an Ensemble of Silicon Vacancies in SiC} 
\author{David A. Fehr}
\email{david-fehr@uiowa.edu}
\affiliation{Department of Physics and Astronomy,  University of Iowa, Iowa City, Iowa 52242, USA}
\author{Hannes Kraus}
\email{hannes.kraus@jpl.nasa.gov}
\affiliation{Jet Propulsion Laboratory, California Institute of Technology, Pasadena, California 91011, USA}
\author{Corey J. Cochrane}
\email{corey.j.cochrane@jpl.nasa.gov}
\affiliation{Jet Propulsion Laboratory, California Institute of Technology, Pasadena, California 91011, USA}
\author{Michael E. Flatt\'{e}}
\email{michaelflatte@quantumsci.net}
\affiliation{Department of Physics and Astronomy,  University of Iowa, Iowa City, Iowa 52242, USA}
\affiliation{Department of Applied Physics, Eindhoven University of Technology, Eindhoven 5612 AZ, The Netherlands}

\date{\today}

\begin{abstract}
We present a quantitative theory for simulating optically detected magnetic resonance (ODMR) measurements of optically-active spin centers using steady-state Lindblad  equations. We apply the theory to an experimental ODMR spectrum associated with the negatively-charged silicon vacancy V2 center in 6H-SiC, showing that spin Hamiltonian parameters, optical transition rates, and even coherence times may be extracted, with values consistent with recent literature. Notably the $T_2$ spin coherence time is measurable, not just the $T_2^*$ dephasing time. Furthermore, we simulate the ODMR spectra of a V2 center in isotopically-purified 6H-SiC, and predict an order-of-magnitude narrowing of some, but not all spectral lines  compared with natural abundance samples. 
\end{abstract}

\maketitle
Color centers in wide bandgap semiconductors, such as the nitrogen-vacancy center in diamond \cite{BalasubramanianGopalakrishnan2009Usct,TaylorJ.M.2008Hdmw,HerbschlebE.D.2019Ucta,Bar-GillN.2013Sesc,DoldeF.2011Esus,PhysRevLett.104.070801,PhysRevLett.112.047601}, the boron vacancy in hexagonal boron nitride \cite{GottschollAndreas2021Sdih,RamsayAndrewJ.2023Cpos,NikhilMathur2022Esso,UdvarhelyiPéter2023Apds,LyuXiaodan2022SQSw}, and the silicon vacancy in silicon carbide \cite{KrausH.2014Mfat,CochraneCoreyJ.2016Vmfs,WidmannMatthias2015Ccos,ChristleDavidJ.2015Iesi,PhysRevApplied.19.044086,PhysRevApplied.20.L031001}, have emerged as key platforms for atomic-scale quantum sensing technologies due to long coherence times \cite{BalasubramanianGopalakrishnan2009Usct,Bar-GillN.2013Sesc,WidmannMatthias2015Ccos,ChristleDavidJ.2015Iesi,RamsayAndrewJ.2023Cpos}, optical accessibility, and susceptibility to fluctuations in local magnetic \cite{BalasubramanianGopalakrishnan2009Usct,HerbschlebE.D.2019Ucta,TaylorJ.M.2008Hdmw,PhysRevApplied.19.044086,KrausH.2014Mfat,CochraneCoreyJ.2016Vmfs,GottschollAndreas2021Sdih} and electric fields \cite{DoldeF.2011Esus,UdvarhelyiPéter2023Apds,PhysRevLett.112.087601,PhysRevLett.112.187601}, temperature \cite{PhysRevLett.104.070801,PhysRevApplied.20.L031001,KrausH.2014Mfat,GottschollAndreas2021Sdih,NikhilMathur2022Esso}, and pressure \cite{PhysRevLett.112.047601,GottschollAndreas2021Sdih,UdvarhelyiPéter2023Apds,LyuXiaodan2022SQSw}. In color center based quantum sensing schemes, optically detected magnetic resonance (ODMR) is the archetypal spectroscopic tool used to read out the response of the color center to the external stimuli. In ODMR, the color center is optically pumped while an applied magnetic field removes the degeneracy of the spin system due to the Zeeman splitting of the energy levels, thus resulting in a photoluminescence (PL) signal that is read out by a detector. RF (typically microwaves) are also applied to induce Rabi oscillations, which cause peaks in the PL contrast when resonant with the energy level splitting. This creates a characteristic and reproducible curve for each color center, as the conditions for magnetic resonance are uniquely determined by the spin Hamiltonian. A detailed and fully quantitative simulation of the ODMR of a color center would provide confidence that the foundations of nonequilibrium optical and spin dynamics relevant for ODMR are sufficiently understood, while allowing for the extraction of reliable parameters for spin coherence times, optical transition rates, and microwave coupling strengths. 

Here we show a quantitative theory for simulating ODMR spectra with Lindblad master equations and that it can be used to extract spin Hamiltonian parameters, optical transition rates, and spin coherence times from experimental spectra. Importantly, our theory allows for the extraction of single-defect $T_{2}$ times from continuous-wave ODMR ensemble measurements, providing a new avenue for $T_{2}$ measurement beyond spin echo experiments. Notably, we provide a full theoretical description of the half-field resonance peaks derive an  estimate of $d_{\perp}$ for silicon vacancies (neither, to our knowledge, have  been known prior). Finally, we predict the effect of isotopic purification on the sharpness of single-defect half-field and two-photon lineshapes, motivating future work which may explore the potential use of these transitions for quantum sensing applications with enhanced sensitivity.

\begin{figure}[ht!]
    
    \centering
    \subfloat[]{
    \includegraphics[width=0.99\linewidth]{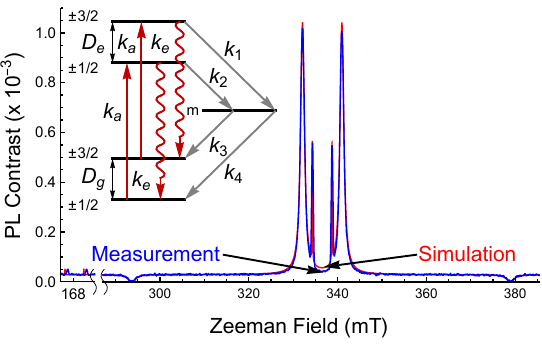}}
    \\
    \subfloat[]{\includegraphics[width=0.99\linewidth]{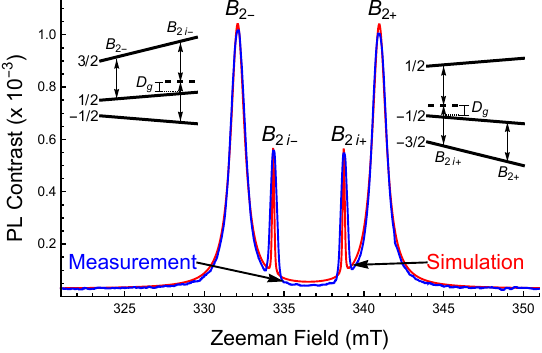}}
  \caption{\justifying(a) ODMR spectrum of the V2 silicon vacancy center in 6H-SiC versus magnetic field with a microwave frequency of 9.4 GHz. The simulated spectra is compared to the measurement in \cite{KrausH.2014Rqme}. (Inset): The ODMR diagram used in our model including the ground and excited state zero-field splittings ($D_{g,e}$), absorption rates ($k_{a}$), spontaneous emission rates ($k_{e}$), and intersystem crossing rates ($k_{1-4}$). (b) ODMR spectrum of the ground resonances $B_{2\pm}$ and $B_{2i\pm}$. (Insets): Energy level diagrams of the transitions associated with $B_{2\pm}$ and $B_{2i\pm}$.}
\label{fig:results no half field}
\end{figure}

We simulate the ODMR spectrum of the negatively-charged silicon vacancy V2 center in 6H-SiC with a 9-level model, with $S=3/2$ ground and excited manifolds separated by 1.397 eV \cite{KrausH.2014Rqme,PhysRevB.98.195204,BreevI.D.2022Ifso,PhysRevX.6.031014,PhysRevB.83.125203,Singh_2023,10.1063/1.5083031,AstakhovG.V.2016SCiS}, and a metastable state in the energy gap between these manifolds. The ODMR diagram of V2 at zero magnetic field is shown in the inset of Fig.~\ref{fig:results no half field}(a). The Hamiltonian for each spin manifold governs the coherent dynamics and the resulting positions of the extrema in the ODMR spectrum. The Hamiltonians of the ground and excited spin manifolds have the following form:
\begin{align}\label{eq:Hamiltonian}
        \hat{H}_{i}&=g_{\parallel,i}\mu_{B}\left(\vec{B}_{0}+\vec{B}_{1}(t)\right)\cdot\hat{\vec{S}}_{i}+D_{i}\left(\hat{S}_{zi}^{2}-\frac{1}{3}S_{i}(S_{i}+1)\right)\notag\\
        &+d_{\perp,i}E_{1}\cos(\omega t)\left\{\hat{S}_{xi},\hat{S}_{yi}\right\}+A_{||,i}\hat{S}_{zi}
\end{align}
where $i\ \epsilon\ \{g,e\}$ for the ground and excited state spins, $\mu_{B}$ is the Bohr magneton, $\vec{S}$ is the vector of $S=3/2$ spin operators, $\vec{B}_{0}$ is the quasi-static  magnetic field taken to be along $\hat{e}_{z}$ $||$ $\hat{c}$, $g_{\parallel}$ is the g-tensor component parallel to $\vec{B}_{0}$, $|\vec{B}_{1}(t)|=B_{1}\cos(\omega t)$ is the magnetic component of the time-dependent microwave field taken to be along $\hat{e}_{x}$ with frequency $\omega$, $D_{i}$ is the axial component of the zero field splitting \cite{KrausH.2014Rqme,PhysRevB.98.195204,BreevI.D.2022Ifso,PhysRevB.93.081207}, $d_{\perp}$ is the perpendicular Stark coupling coefficient \cite{PhysRevLett.112.087601,PhysRevLett.112.187601,PhysRevB.110.024419}, and $E_{1}$ is the magnitude of an in-plane time-dependent electric field taken to be along $\hat{e}_{y}$. In this case, $E_{1}$ is the electric component of the applied microwaves, with strength $E_{1}=cB_{1}$. Finally, $A_{||,i}$ is the component of an effective magnetic field distribution parallel to the quasi-static magnetic field which models both the hyperfine splitting of $^{29}$Si (4.7$\%$ abundance) and $^{13}$C nuclei (1.1$\%$ abundance), and ensemble-related magnetic inhomogeneities.

\begin{figure}[ht!]
    
    \centering
    \subfloat[]{\includegraphics[width=0.99\linewidth]{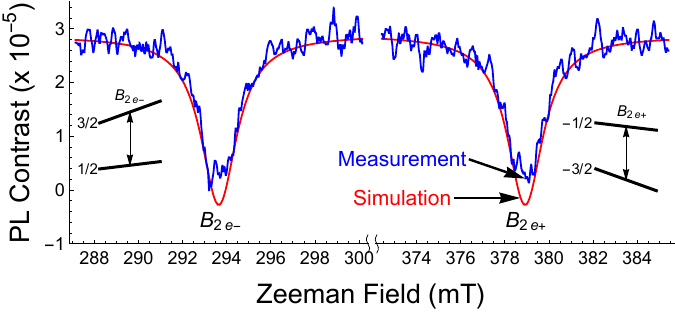}
    }
    \\
    \subfloat[]{\includegraphics[width=0.99\linewidth]{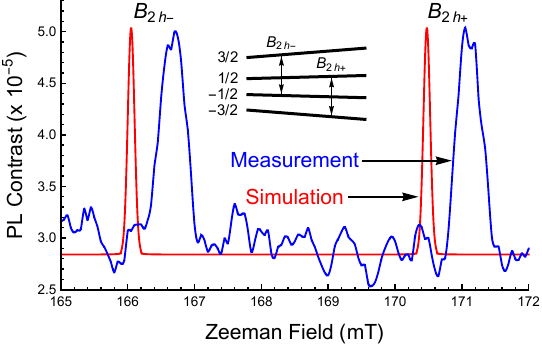}
    }
  \caption{\justifying(a) ODMR spectrum of the excited state resonances $B_{2e\pm}$. (Insets): Energy level diagrams of the transitions associated with $B_{2e\pm}$. (a) ODMR spectrum of the ground state half-field resonances $B_{2h\pm}$. (Insets): Energy level diagrams of the transitions associated with $B_{2h\pm}$.}
\label{fig:results-excited&half field}
\end{figure}

PL contrast measurements of V2 at 4 Kelvin in 6H-SiC from \cite{KrausH.2014Rqme} and corresponding simulations from our theory are shown in Figs.~\ref{fig:results no half field} and \ref{fig:results-excited&half field} as functions of the quasi-static (dc) magnetic field and at a constant microwave frequency of $f=\omega/2\pi\simeq$ 9.4 GHz. Figure~\ref{fig:results no half field}a shows the full ODMR spectra of V2, with measurement in blue and simulation in red, while the inset explicitly shows the ODMR level structure and rates used in our model. Figure~\ref{fig:results no half field}b zooms in on the central feature in Fig.~\ref{fig:results no half field}a and more clearly shows the comparison between our simulation and the measurement for ground resonances $B_{2\pm}$ and $B_{2i\pm}$, with labeling consistent with the notation in \cite{KrausH.2014Rqme}. $\mathrm{B}_{\mathrm{2\pm}}$ correspond to single-photon $\Delta m=1$ transitions between the $|\pm3/2\rangle\leftrightarrow|\pm1/2\rangle$ ground spin states. $\mathrm{B}_{\mathrm{2i\pm}}$ correspond to two-photon $\Delta m=2$ transitions between the $|\pm3/2\rangle\leftrightarrow|\mp1/2\rangle$ ground spin states. The two-photon resonances can be explained as two-photon transitions, with occupation of a virtual intermediate state detuned from $|\pm1/2\rangle$ by an amount equal to the ground state zero-field splitting ($D_{g}$). The energy level diagrams for these transitions are shown in the insets in Fig.~\ref{fig:results no half field}b. The difference and sum of the $B_{2\pm}$ peak positions, $\Delta(\Sigma) B_{2\pm}\equiv B_{2+}-(+)B_{2-}$, can be used to determine $D_{g}$ and $g_{g}$ as $D_g=g_{g}\mu_{B}\Delta B_{2\pm}/4h$ and $g_{g}=2hf/\mu_{B}\Sigma B_{2\pm}$. We find $2D_{g}/h=124$ MHz, which closely agrees with \cite{KrausH.2014Rqme,PhysRevB.98.195204,BreevI.D.2022Ifso,PhysRevB.61.2613,10.1063/1.5083031,Singh_2023,AstakhovG.V.2016SCiS}, and $g_{g}=2.003$, which closely agrees with \cite{KrausH.2014Rqme,PhysRevB.98.195204,BARANOV2001680,PhysRevX.6.031014,PhysRevB.61.2613,AstakhovG.V.2016SCiS}. Together, these values of $D_{g}$ and $g_{g}$ uniquely identify this defect as V2 in 6H-SiC.

The resonances in Fig.~\ref{fig:results no half field}b also provide information about the ground state coherence times $T_{1,g}$ and $T_{2,g}$, inhomogeneous dephasing time $T_{2,g}^{*}$, microwave magnetic field strength $B_{1}$, and optical transition rates $k_{a}$ and $k_{e}$. At low temperature, the linewidths of $B_{2\pm}$ are primarily microwave power broadened and we extract $B_{1}=0.25$ mT in Eq.\:\eqref{eq:Hamiltonian}. Homogeneous spin dephasing also contributes to these linewidths, however minutely at low temperature, and we estimate the lower bound of the ground spin dephasing time to be $T_{2,g}\geq$ 29 $\mu$s. To our knowledge, a 4 Kelvin ODMR measurement of $T_{2,g}$ for V2 in 6H-SiC has not been published. However, similar measurements have been published which provide perspective on our extracted lower bound. 6H-SiC measurements at room temperature in \cite{PhysRevB.95.045206} found $T_{2,g}=3.31$ $\mu$s for V2, and $T_{2,g}=3.37$ $\mu$s for V1 and V3. On the other hand, 4H-SiC measurements at 4 Kelvin in \cite{NagyRoland2019Hsao} found $T_{2,g}=850$ $\mu$s for V1. Therefore, we find $T_{2,g}\geq$ 29 $\mu$s to be a reasonable lower bound for the measurement analyzed from \cite{KrausH.2014Rqme}. 

Next, we focus on the lineshapes of $B_{2i\pm}$. While the amplitudes are strongly affected by $B_{1}$, the linewidths respond more weakly to homogeneous dephasing than $B_{2\pm}$. Interestingly, we found $B_{2i\pm}$ responds strongly to inhomogeneous broadening and could be a quantitative measure of $T_{2,g}^{*}$. We convolved our simulation with a Gaussian distribution as in Eq.\:\eqref{eq:PL Contrast Avg} to account for magnetic inhomogeneities (hyperfine fields, spatial and temporal fluctuations of $g_{g}$ and $B_{0}$, etc.) and fit the standard deviation. We found $A_{\sigma}$ = 1.4 MHz. $A_{\sigma}$ is also related to $T_{2}^{*}$ as $T_{2}^{*}\simeq1/\pi A_{\sigma}$\cite{PhysRevB.103.195201}, and we estimate $T_{2}^{*}\simeq220$ ns, in order-of-magnitude agreement with recent measurements \cite{PhysRevB.101.134110,PRXQuantum.3.010343,PhysRevB.95.161201}. We also found the amplitudes of $B_{2i\pm}$ were strongly attenuated by $T_{1,g}$, which allowed us to constrain $T_{1,g}\geq$ 2 ms. For perspective, recent room temperature measurements found $T_{1,g}=107$ $\mu$s \cite{NagyRoland2019Hsao}, while the upper bound of $T_{1,g}$ in 4H-SiC has been predicted to be on the order of seconds at cryogenic temperatures \cite{PhysRevB.95.161201}. Finally, we fit the absorption ($k_{a}$) and spontaneous emission rates ($k_{e}$) shown in the inset of Fig.\:\ref{fig:results no half field} to match the absolute amplitudes of the $B_{2\pm}$ and $B_{2i\pm}$ resonances with the measurement in \cite{KrausH.2014Rqme}. We found $1/k_{a}=2.5$ $\mu$s and $1/k_{e}=105$ ns. While $k_{a}$ is an experiment-dependent parameter proportional to the optical power density, there are reported values of $1/k_{e}\sim 10$ ns in the literature for V2 in 4H-SiC  \cite{PhysRevApplied.11.024013,LiuDi2024Tsvc}, which both assume a more complicated level structure than considered in this work. We found it was not possible to find a good fit with the $\sim10\times$ faster spontaneous emission rate of V2 in 4H-SiC due to significant broadening of $B_{2\pm}$.

Fig.~\ref{fig:results-excited&half field}a zooms in on the small minima in Fig.~\ref{fig:results no half field}a (near approximately 294 mT and 379 mT) and more clearly shows the comparison between our ODMR simulation and the measurement from \cite{KrausH.2014Rqme} for excited resonances $B_{2e\pm}$. These resonances correspond to single-photon $\Delta m=1$ transitions between the $|\pm3/2\rangle\leftrightarrow|\pm1/2\rangle$ excited states, and are analogous to $\mathrm{B}_{\mathrm{2\pm}}$. From the peak positions of $B_{2e\pm}$ we may extract $D_{e}$ and $g_{e}$ in the same way as for $D_{g}$ and $g_{g}$. We find $2D_{e}/h=1.22$ GHz in agreement with \cite{BreevI.D.2022Ifso,Singh_2023,AstakhovG.V.2016SCiS}, and $g_{e}=2.0045$. We fit the linewidths of $B_{2e\pm}$ by including homogeneous spin dephasing for the excited state, extracting $T_{2,e}$ = 16 ns in agreement with the estimate of 10 ns in \cite{KrausH.2014Rqme}. Finally, we estimated the lower bound of the spin relaxation time $T_{1,e}\geq$ = 10 $\mu$s by fitting the absolute amplitude of $B_{2e\pm}$ in the simulation to the measurement.

One crucial step in simulating ODMR with the method we present here is scaling the absolute amplitudes of the ground and excited resonances relative to each other to match the experimental measurement, such as in \cite{KrausH.2014Rqme}. We found that the nonradiative intersystem crossing rates, $k_{1-4}$ in Fig.~\ref{fig:results no half field}a, to be the key to this scaling. We find $k_{1}$ = 31.4 $\mu s^{-1}$, $k_{2}$ = 27.8 $\mu s^{-1}$, $k_{3}$ = 1.8 $\mu s^{-1}$, and $k_{4}$ = 1.6 $\mu s^{-1}$. Similar numbers have been published for models of higher complexity for V2 in 4H-SiC \cite{PhysRevApplied.11.024013,LiuDi2024Tsvc}.

Fig.~\ref{fig:results-excited&half field}b zooms in on the smallest maxima in Fig.~\ref{fig:results no half field}a (near approximately 166 mT and 171 mT) and more clearly shows the comparison between our ODMR simulation and the measurement from \cite{KrausH.2014Rqme} for the half-field resonances $B_{2h\pm}$. These resonances correspond to single-photon $\Delta m=2$ transitions between the $|\pm\frac{3}{2}\rangle\leftrightarrow|\mp\frac{1}{2}\rangle$ states in the ground manifold, mediated by the oscillating electric field component of the microwave. Spin-orbit coupling mixes the orbital and spin angular momenta \cite{PhysRevB.93.081207} and, with V2 lacking inversion symmetry from the $C_{3v}$ site symmetry, the spin states can then be electrically manipulated \cite{PhysRevLett.112.087601,PhysRevLett.112.187601,PhysRevB.110.024419}. The coupling strength to in-plane electric fields is denoted by the parameter $d_{\perp}$ in Eq.\:\eqref{eq:Hamiltonian}. Using $E_{1}=cB_{1}$ = 750 $\mathrm{V}/\mathrm{cm}$, we estimate $d_{\perp}/{h}\approxeq$ 45 Hz$/\mathrm{V}\cdot\mathrm{cm}^{-1}$. For comparison, $d_{\perp}/{h}=$ 17, 26 Hz$/\mathrm{V}\cdot\mathrm{cm}^{-1}$ for the NV${}^{-}$ center in diamond and QL1 in 6H-SiC, respectively \cite{PhysRevLett.112.087601,PhysRevLett.112.187601}, both of which have $C_{3v}$ symmetry. Furthermore, $d_{\perp}/{h}=$ 32.3, 28.5, 32.5 Hz$/\mathrm{V}\cdot\mathrm{cm}^{-1}$ for $C_{1h}$-symmetric divacancies in 4H-SiC (PL3, PL4, and PL5, respectively) \cite{PhysRevLett.112.187601}. Therefore, we consider the value of $d_{\perp}/{h}\approxeq$ 45 Hz$/\mathrm{V}\cdot\mathrm{cm}^{-1}$ which we fit to the ODMR spectra of V2 in 6H-SiC to be reasonable since it is on a similar order to other color centers in silicon carbide and diamond. Lastly, we point out the presence of a uniform shift ($\approx0.63$ mT) in the measured ODMR spectrum from \cite{KrausH.2014Rqme} which our simulations do not naturally reproduce. This shift is within the uncertainty (up to 1 mT) in the quasi-static magnetic field source used near $B_{2h\pm}$, a Hall probe, in contrast to the uncertainty in the NMR probe ($\sim2$ $\mu$T) used near $B_{2\pm}$, $B_{2i\pm}$, and $B_{2e\pm}$.  Another possible explanation could be a misalignment in the quasi-static magnetic field source near $B_{2h\pm}$. Using the ground-state Hamiltonian from Eq.\:\eqref{eq:Hamiltonian}, we may explicitly include the dependence of the peak positions of $B_{2h\pm}$ on the polar angle of $\vec{B}_{0}$ measured from the c-axis:
\begin{align}
    B_{2h\pm}\cos(\theta)=\frac{(f\pm2D_{g})h}{2g_{g}\mu_{B}}
    \label{eq:half-field calibration}
\end{align} 
where the theoretical values of the resonances in Eq.\:\eqref{eq:half-field calibration} correspond to $\theta=0\degree$, and the experimental values correspond to $\theta\neq0\degree$. We also point out that Eq.\:\eqref{eq:half-field calibration} effectively defines a calibration curve, allowing one to align the sample and magnetic field source self consistently. Using the parameters $f=9.43451$ GHz, $D_{g}/h=62.20157$ MHz, $g_{g}=2.0030023$, and $\Delta \mathrm{B}_{\mathrm{2h\pm}}=0.62675$ mT, where $\Delta \mathrm{B}_{\mathrm{2h\pm}}$ is the difference between the experimental and theoretical peak positions, we estimate a misalignment of $\theta=4.97973\degree$. 

\begin{figure}[ht!]
    
    \centering
    \subfloat[]{\includegraphics[width=0.99\linewidth]{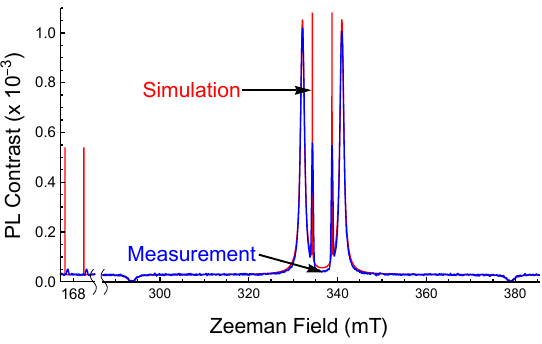}}
    \\
    \subfloat[]{\includegraphics[width=0.99\linewidth]{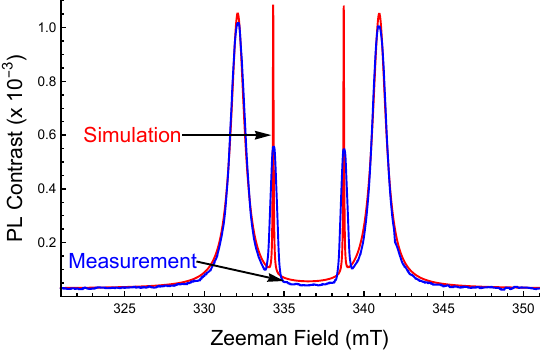}}
    \\
    \subfloat[]{\includegraphics[width=0.99\linewidth]{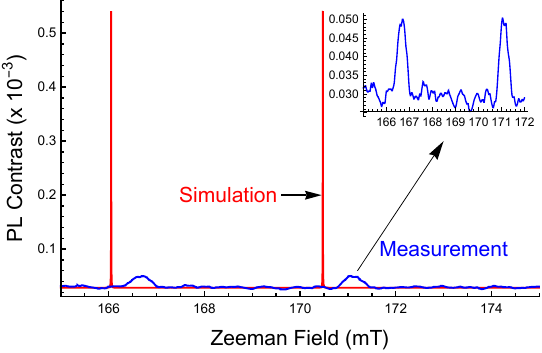}}
  \caption{\justifying(a) The predicted ODMR spectrum of the V2 silicon vacancy center in isotopically-purified and magnetically-homogeneous 6H-SiC versus magnetic field with a microwave frequency of 9.4 GHz. The simulated spectra is compared to the natural abundance measurement in \cite{KrausH.2014Rqme} for reference. (b) Predicted ODMR spectrum of the ground resonances $B_{2\pm}$ and $B_{2i\pm}$. (b) Predicted ODMR spectrum of the ground state half-field resonances $B_{2h\pm}$.}
\label{fig:3}
\end{figure}

We also investigated how the ODMR spectrum might change for the case of a single V2 color center in a perfectly isotopically-purified sample at 4 K, and if this would be advantageous for quantum sensing applications. In our simulations this amounts to taking the limit of $A_{\sigma}\rightarrow0$ in Eq.\:\eqref{eq:hyperfine integral}. The results of these simulations are shown in Fig.~\ref{fig:3}(a)-(c). We see in Fig.~\ref{fig:3}(a) that the $B_{2\pm}$ and $B_{2e\pm}$ resonances are not significantly affected by removing magnetic inhomogeneities, suggesting that these resonances are rather robust to inhomogeneous broadening ($T_{2,g}^{*}$). In contrast, we see in Fig.~\ref{fig:3}(b)-(c) that the $B_{2i\pm}$ and $B_{2h\pm}$ resonances are dramatically affected by an isotopically-pure and magnetically-homogeneous environment, suggesting an extreme susceptibility to inhomogeneous broadening. The most remarkable differences can be seen in the half-field resonances, which exhibit an amplification of $\simeq23\times$ in the PL contrast  and an attenuation of $\simeq34\times$ in the full width at half maximum. In a simple picture one might expect this to correspond to an order of magnitude improvement in sensitivity for $B_{2i\pm}$ and $B_{2h\pm}$, as sensitivity $\eta\propto\Delta\nu/C$, where $\Delta\nu$ is the linewidth and $C$ is the PL contrast \cite{PhysRevB.84.195204,BarryJohnF.2016Omdo}.

We now lay out our generalized theoretical framework for simulating ODMR of color centers. We model the combination of optical and spin dynamics present in an ODMR measurement with a Lindblad master equation \cite{GoriniVittorio1976Cpds,LindbladG.1976Otgo,10.1093/acprof:oso/9780199213900.001.0001} of the form: 
\begin{align}\label{eq:Master Equation}
       \partial_{t}\hat{\rho}(t)&=-\frac{i}{\hbar}\left[\hat{H}(t),\hat{\rho}(t)\right]+\sum_{i}\hat{\mathcal{L}}_{i}\left[\hat{\rho}(t)\right]\notag\\ \hat{\mathcal{L}}_{i}\left[\hat{\rho}(t)\right]&\equiv k_{i}\left(\hat{L}_{i}\hat{\rho}(t)\hat{L}_{i}^{\dagger}-\frac{1}{2}\left\{\hat{L}_{i}^{\dagger}\hat{L}_{i},\hat{\rho}(t)\right\}\right)
\end{align}
 where $\hat{H}(t)$ is block diagonal in the Hamiltonians of the individual manifolds defined in Eq.\:\eqref{eq:Hamiltonian}. The first term on the right side of Eq.\:\eqref{eq:Master Equation} determines the coherent evolution of the spin states. The second term, $\hat{\mathcal{L}}_{i}\left[\hat{\rho}(t)\right]$, models the optical and intersystem crossing transitions with the following Lindblad operators:
\begin{align}\label{eq:optical cycle operators}
   \begin{matrix}
       \hat{L}_{a,m_{s}}=|e,m_{s}\rangle\langle g,m_{s}|&&\hat{L}_{e,m_{s}}=|g,m_{s}\rangle\langle e,m_{s}|\\
       \hat{L}_{e,m_{s}\rightarrow m}=|m\rangle\langle e,m_{s}|&&
    \hat{L}_{m\rightarrow g,m_{s}}=|g,m_{s}\rangle\langle m|
   \end{matrix}
\end{align}
 where g, e, and m refer to ground, excited, and metastable states, respectively, while $m_{s}$ refers to the spin quantum number. Each of the operators in Eq.\:\eqref{eq:optical cycle operators} have associated rates $k_{a,m_{s}}$, $k_{e,m_{s}}$, $k_{e,m_{s}\rightarrow m}$, and $k_{m\rightarrow g,m_{s}}$. For the simulations of V2 ODMR we use $k_{a,m_{s}}\equiv k_{a}$, $k_{e,m_{s}}\equiv k_{e}$, $k_{e,\pm3/2\rightarrow m}\equiv k_{1}$, $k_{e,\pm1/2\rightarrow m}\equiv k_{2}$, $k_{m\rightarrow g,\pm3/2}\equiv k_{3}$, and $k_{m\rightarrow g,\pm1/2}\equiv k_{4}$ in Fig.~\ref{fig:results no half field}(a).
 
 We also use $\hat{\mathcal{L}}_{i}\left[\hat{\rho}(t)\right]$ in Eq.\:\eqref{eq:Master Equation} to model the effects of finite coherence times with Lindblad operators \cite{PhysRevB.110.024419}:
 \begin{align}\label{eq:coherence operators}
     &\begin{matrix}
     \hat{L}_{r,m_{s}\leftrightarrow m_{s}^{\prime}}=|m_{s}\rangle\langle m_{s}^{\prime}|&&  \hat{L}_{r,m_{s}^{\prime}\leftrightarrow m_{s}}=\hat{L}_{r,m_{s}\leftrightarrow m_{s}^{\prime}}^{\dagger}\notag
     \end{matrix}\\
 &\hat{L}_{d,m_{s}\leftrightarrow m_{s}^{\prime}}=|m_{s}\rangle\langle m_{s}|-|m_{s}^{\prime}\rangle\langle m_{s}^{\prime}|
 \end{align}
 where $\hat{L}_{r,m_{s}\leftrightarrow m_{s}^{\prime}}$ and $\hat{L}_{r,m_{s}^{\prime}\leftrightarrow m_{s}}$ are $\hat{\sigma}_{\pm}$ in the subspace spanned by $\left\{|m_{s}\rangle,|m_{s}^{\prime}\rangle\right\}$ and model spin relaxation at a rate $k_{r,m_{s}\leftrightarrow m_{s}^{\prime}}$, while $\hat{L}_{d,m_{s}\leftrightarrow m_{s}^{\prime}}$ is $\hat{\sigma}_{z}$ and models (homogeneous) spin dephasing of the subspace at a rate $k_{d,m_{s}\leftrightarrow m_{s}^{\prime}}$. For the simulations of V2 ODMR we use a single relaxation rate $k_{r}\equiv k_{r,m_{s}\leftrightarrow m_{s}^{\prime}}$ and dephasing rate $k_{d}\equiv k_{d,m_{s}\leftrightarrow m_{s}^{\prime}}$ for all $m_{s}\neq m_{s}^{\prime}$ subspaces. We then construct the coherence times and constraint for $S=3/2$:
 \begin{align}\label{eq:coherence times}
     &\begin{matrix}
         T_{1}=\left(4k_{r}\right)^{-1}&&T_{2}=\left(3k_{r}+4k_{d}\right)^{-1}\notag\\
     \end{matrix}\\
     &3T_{2}\leq4T_{1}
 \end{align}
 where the constraint between $T_{1}$ and $T_{2}$ in Eq.\:\eqref{eq:coherence times} is derived by requiring $k_{d}\geq0$. A different construction is necessary to distinguish between the coherence times of the $\Delta m=1$ and $\Delta m=2$ transitions, and this elucidation is the subject of future work.

The steady-state PL is determined by the density matrix populations of the excited state and the rate(s) of spontaneous emission. Thus, we define the PL operator, PL, and PL contrast:
\begin{align}
    \widehat{PL}&\equiv\sum_{m_{s}}k_{e,m_{s}}\hat{L}^{\dagger}_{e,m_{s}}\hat{L}_{e,m_{s}}\label{eq:PL operator}\\
    PL(B_{1},B_{0})&\equiv Tr\left[\widehat{PL}\hat{\rho}(\tau_{ss})\right]\label{eq:PL}\\
    \frac{\Delta PL}{PL}\left(B_{0}\right)&\equiv\frac{PL(B_{1},B_{0})-PL(0,B_{0})}{PL(0,B_{0})}\label{eq:PL Contrast}
\end{align}
where the density matrix in Eq.\:\eqref{eq:PL} is the solution of Eq.\:\eqref{eq:Master Equation} in the steady-state $\partial\rho(\tau_{ss})/\partial t\equiv0$. Finally, inhomogeneous effects are treated by assigning a normal Gaussian distribution to any quantities which vary across the ensemble (e.g. $g_{\parallel}$, $D$, $k_{i}$, etc.) and integrating over the spectrum:
\begin{align}
    \frac{\Delta PL}{PL}\left(B_{0}\right)&\equiv\int_{-\infty}^{+\infty}P(x,\sigma)\frac{\Delta PL}{PL}\left(B_{0},x\right)\,dx\label{eq:PL Contrast Avg}\\
    P(x,\sigma)&\equiv\sqrt{\frac{1}{2\pi \sigma^{2}}}e^{-\frac{1}{2}\left(\frac{x-\mu}{\sigma}\right)^{2}}\label{eq:hyperfine integral}
\end{align}
where for the treatment of magnetic inhomogeneities in the ODMR simulations of V2: $A_{\parallel}\equiv x$, $A_{\sigma}\equiv\sigma=1.4$ MHz, and $A_{\mu}\equiv\mu=0$ MHz. For ease of calculation, we truncated the integral in Eq.\:\eqref{eq:PL Contrast Avg} at $\mu\pm3\sigma=\pm4.2$ MHz, which captures 99.7\% of the total spectrum.

We now discuss the impact of this theory for quantitative simulations of spin-coherent phenomena with steady-state Lindblad equations. This work exhibits our theory for simulating the ODMR spectrum of the V2 ($S=3/2$) color center in 6H-SiC and, by fitting our simulations to the measured spectrum from \cite{KrausH.2014Rqme}, we extracted reasonable values of spin Hamiltonian parameters, optical transition rates, and spin coherence times. This model can be immediately applied to the other silicon vacancy centers in 6H and 4H-SiC by including the appropriate parameters in Eq.\:\eqref{eq:Hamiltonian} and Eq.\:\eqref{eq:Master Equation}. With minor modifications to the ODMR diagram and level structure, our theory could also be extended to simulate ODMR spectra of color centers with $S\neq3/2$ such as the nitrogen vacancy in diamond, boron vacancy in hBN, or divacancies in SiC (all $S=1$), widening the applicability of our theory beyond magnetometry to temperature, pressure, and electric field sensing. Furthermore, as the Lindblad operators in Eq.\:\eqref{eq:optical cycle operators} merely act to incoherently couple the different spin manifolds, more complicated models of ODMR spectroscopy may be considered such as nontrivial metastable manifold dynamics \cite{PhysRevApplied.11.024013,LiuDi2024Tsvc,PatelRajN.2024RTDo}, coupled spin pairs \cite{gao2024singlenuclearspindetection,robertson2025chargetransfermechanismoptically,singh2024violetnearinfraredopticaladdressing}, transition metal \cite{BosmaTom2018Iato,PhysRevApplied.22.044078,DilerBerk2020Ccah}, molecular \cite{PhysRevLett.133.120801,PhysRevX.12.031028}, and rare earth spin-optical dynamics \cite{PhysRevA.103.022618,SerranoD.2018Acol}. Finally we point out that the success of steady-state Lindblad equations for the quantitative simulation of ODMR spectroscopy opens the door to simulating related electrical spectroscopies of spin centers such as electrically detected magnetic resonance, near-zero field magnetoresistance, or spin-polarized transport as an alternative to non-equilibrium Green's functions \cite{PhysRevLett.113.047205,PhysRevB.76.045213} and stochastic Liouville formalisms \cite{PhysRevLett.125.257203,9039723}.

\begin{acknowledgments}
AFOSR FA9550-22-1-0308 supported  study of the silicon vacancy as a radiation-induced defect in a wide bandgap semiconductor and its identification with ODMR. NSF DMR-1921877 supported the development of the ODMR simulation code. An Iowa NASA EPSCoR Seed Grant initiated the study of magnetic field sensitivity.  The JPL Visiting Student Research Program supported initial selection of the problem during a summer visit of D. A. F. to JPL. We wish to thank V. Dyakonov, N. J. Harmon, S. R. McMillan and A. Sperlich for valuable discussions, and V. D. and A. S. also for providing the experimental data files from \cite{KrausH.2014Rqme}. A portion of this research was carried out at the Jet Propulsion Laboratory, California Institute of Technology, under contract with the National Aeronautics and Space Administration, (contract 80NM0018D0004).
  
\end{acknowledgments}


\bibliography{references}
\bibliographystyle{apsrev4-2}


\end{document}